\documentstyle[aps,eqsecnum,floats]{revtex}

\begin{document}
\draft
\twocolumn[\hsize\textwidth\columnwidth\hsize\csname
           @twocolumnfalse\endcsname

\title{Symmetric hyperbolic system in the Ashtekar formulation
\cite{status}}
\author{Gen Yoneda \cite{Email-yone}}
\address{
Department of Mathematical Science, Waseda University,
Okubo 3-4-1, Shinjuku, Tokyo 169-8790, Japan}
\author{Hisa-aki Shinkai  \cite{Email-his}}
\address{
Department of Physics, Washington University,
St. Louis, MO 63130-4899, USA
}
\date{submitted March 24, 1998; ~~ revised Sep. 23, 1998}
\maketitle
\begin{abstract}
\widetext

We present a first-order symmetric hyperbolic system
in the Ashtekar formulation of general relativity for vacuum 
spacetime.  We add terms from the constraint equations to
 the evolution equations
with appropriate combinations, which is the same technique used by
Iriondo, Leguizam\'on and Reula  ( Phys. Rev. Lett. 79, 4732 (1997) ).
However our system is different from theirs
in the points that we  primarily use 
Hermiticity of a characteristic matrix of the system 
to characterize our system
{\it symmetric},
discuss the consistency of this system with reality condition,
and show the characteristic speeds of the system.
\end{abstract}

\pacs{PACS numbers: 04.20.Cv, 04.20.Ex, 04.20.Fy}

\vskip 2pc]
\narrowtext



\section{Introduction} \label{sec:intro}

Hyperbolic formulation of the Einstein equation is the one of the 
main research areas in general relativity \cite{Reula98}.
This formulation is used in the proof of the existence, uniqueness 
and stability (well-posedness)
of the solutions of the Einstein equation by analytical methods
\cite{Heldbook}.
So far, several first order hyperbolic formulations are proposed; 
some of them
are flux conservative \cite{BonaMasso},
symmetrizable \cite{ChoquetBruhatYork},
or
symmetric hyperbolic system
\cite{FischerMarsden72,fried96,FrittelliReula96}.
The recent interest in hyperbolic formulation
 arises from its application to numerical relativity.
One of the expected advantages is the existence of the
characteristic speeds of the
system, with which we may treat the numerical boundary with appropriate
condition.
Some numerical tests have been reported along this direction
\cite{miguel,SBCSThyper,cactus1}.

Recently, Iriondo, Leguizam\'on and Reula (ILR) \cite{Iriondo}
discuss a symmetric
hyperbolic system in the Ashtekar formulation \cite{Ashtekar}
 of general relativity.
Ashtekar's formulation has many
advantages in the treatment of gravity.
By using his special pair of variables,
the constraint equations which appear in the theory become
low-order polynomials, and the theory has the
correct form for gauge theoretical features. These
suggest possibilities for treating a quantum
description of gravity nonperturbatively.
The classical applications of the Ashtekar's formulation has also 
been discussed by several authors. For 
example, we \cite{ys-con} discussed
the reality conditions for metric and triad and proposed new
set of variables from the point of Lorentzian dynamics. 
We \cite{ysn-dege}
also showed an example of passing degenerate
point in 3-space by loosing reality condition locally.

In this {\it Letter}, we present a new symmetric hyperbolic system
in the Ashtekar's formulation for Lorentzian vacuum
spacetime.
ILR \cite{Iriondo} says that they construct a
symmetric hyperbolic system.
However, we think their discussion is not clear in the 
the following three points.
First,
they used {\it anti}-Hermiticity of 
the pricipal symbol for defining their system {\it symmetric}.
We, however, think that this does not derive Hermiticity of
the characteristic matrix [$A$ below in eq. (\ref{def}),
$B$ in \cite{Iriondo}], 
since they do not define their vector $k_a$ explicitly. 
We rather use 
Hermiticity of the characteristic matrix primarily
to construct a {\it symmetric hyperbolic} system.
Second, they did not mention the consistency of their formulation 
with the reality conditions which are
crucial in the study of the Lorentzian dynamics in the Ashtekar
variables\cite{IRLsecond}.
Third,
they did not discuss the characteristic structure of the system,
which should be shown in the normal hyperbolic formulations.
Our discussion covers these two matters.

The construction of this paper is as follows.
After giving a brief review of Ashtekar's variables and  reality
conditions in \S 2, we present our formulation in \S 3. The
discussion of characteristic speed and summary are
in \S 4.

\section{Ashtekar's formulation}\label{sec:ash}

The key feature of  Ashtekar's formulation of general relativity
\cite{Ashtekar} is the introduction of a self-dual
connection as one of the basic dynamical variables.
Let us write\footnote{
We use $\mu,\nu=0,\cdots,3$ and
$i,j=1,\cdots,3$ as spacetime indices, while
$I,J=(0),\cdots,(3)$ and
$a,b=(1),\cdots,(3)$ are $SO(1,3)$, $SO(3)$ indices respectively.
We raise and lower
$\mu,\nu,\cdots$ by $g^{\mu\nu}$ and $g_{\mu\nu}$ 
(Lorentzian metric);
$I,J,\cdots$ by $\eta^{IJ}={\rm diag}(-1,1,1,1)$ and $\eta_{IJ}$;
$i,j,\cdots$ by $\gamma^{ij}$ and $\gamma_{ij}$(3-metric).
We use volume forms $\epsilon_{abc}$;
$\epsilon_{abc} \epsilon^{abc}=3!$.
}
the metric $g_{\mu\nu}$ using the tetrad, $e^I_\mu$, and
define its inverse, $E^\mu_I$, by
$g_{\mu\nu}=e^I_\mu e^J_\nu \eta_{IJ}$ and
$E^\mu_I:=e^J_\nu g^{\mu\nu}\eta_{IJ}$.
We define SO(3,C) self-dual and anti self-dual 
connections
$~^{\pm\!}{\cal A}^a_{\mu}
:= \omega^{0a}_\mu \mp ({i / 2}) \epsilon^a_{~bc}\omega^{bc}_\mu$,
where $\omega^{IJ}_{\mu}$ is a spin connection 1-form (Ricci
connection), $\omega^{IJ}_{\mu}:=E^{I\nu} \nabla_\mu e^J_\nu.$
Ashtekar's plan is to use  only a self-dual part of
the connection
$^{+\!}{\cal A}^a_\mu$
and to use its spatial part $^{+\!}{\cal A}^a_i$
as a dynamical variable.
Hereafter,
we simply denote $^{+\!}{\cal A}^a_\mu$ as ${\cal A}^a_\mu$.

The lapse function, $N$, and shift vector, $N^i$,
are expressed as $E^\mu_0=(1/N, -N^i/N$).
This allows us to think of
$E^\mu_0$ as a normal vector field to $\Sigma$
spanned by the condition $t=x^0=$const.,
which plays the same role as that of ADM.
Ashtekar  treated the set  (${\cal A}^a_{i}$, $\tilde{E}^i_{a}$)
as basic dynamical variables, where
$\tilde{E}^i_{a}$ is an inverse of the densitized triad
defined by
$\tilde{E}^i_{a}:=e E^i_{a}$
where $e:=\det e^a_i$ is a density.
This pair forms the canonical set.

In the case of pure gravitational spacetime,
the Hilbert action takes the form
\begin{eqnarray}
S&=&\int {\rm d}^4 x
[ (\partial_t{\cal A}^a_{i}) \tilde{E}^i_{a}
+(i/2) \null \! \mathop {\vphantom {N}\smash N}
\limits ^{}_{^\sim}\!\null \tilde{E}^i_a
\tilde{E}^j_b F_{ij}^{c} \epsilon^{ab}_{~~c}
- \Lambda \null \! \mathop {\vphantom {N}\smash N}
\limits ^{}_{^\sim}\!\null \det\tilde{E} \nonumber \\ &&
-N^i F^a_{ij} \tilde{E}^j_a
+{\cal A}^a_{0}{\cal D}_i \tilde{E}^i_{a} ],
 \label{action}
\end{eqnarray}
where $\null \! \mathop {\vphantom {N}\smash N}
\limits ^{}_{^\sim}\!\null := e^{-1}N$, $\Lambda$
is the cosmological constant,
${\cal D}_i \tilde{E}^i_{a}
    :=\partial_i \tilde{E}^i_{a}
-i \epsilon_{ab}^{~~c}{\cal A}^b_{i}\tilde{E}^i_{c}$, and
${\rm det}\tilde{E}$ is defined to be
${\rm det}\tilde{E}=
(1/6)\epsilon^{abc}
\null\!\mathop{\vphantom {\epsilon}\smash \epsilon}
\limits ^{}_{^\sim}\!\null_{ijk}\tilde{E}^i_a \tilde{E}^j_b
\tilde{E}^k_c$, where
$\epsilon_{ijk}:=\epsilon_{abc}e^a_ie^b_je^c_k$
 and $\null\!\mathop{\vphantom {\epsilon}\smash \epsilon}
\limits ^{}_{^\sim}\!\null_{ijk}:=e^{-1}\epsilon_{ijk}$
\footnote{$\epsilon_{xyz}=e$,
$\null\!\mathop{\vphantom {\epsilon}\smash \epsilon}
\limits ^{}_{^\sim}\!\null_{xyz}=1$,
$\epsilon^{xyz}=e^{-1}$,
$\tilde{\epsilon}^{xyz}=1$. }.

Varying the action with respect to the non-dynamical variables
$\null \!
\mathop {\vphantom {N}\smash N}\limits ^{}_{^\sim}\!\null$,
$N^i$
and ${\cal A}^a_{0}$ yields the constraint equations,
\begin{eqnarray}
{\cal C}_{H} &:=&
 (i/2)\epsilon^{ab}_{~~c}
\tilde{E}^i_{a} \tilde{E}^j_{b} F_{ij}^{c}
  -\Lambda \det\tilde{E}
   \approx 0, \label{c-ham} \\
{\cal C}_{M i} &:=&
  -F^a_{ij} \tilde{E}^j_{a} \approx 0, \label{c-mom}\\
{\cal C}_{Ga} &:=&  {\cal D}_i \tilde{E}^i_{a}
 \approx 0,  \label{c-g}
\end{eqnarray}
where ${F}^a_{\mu\nu} := (d {\cal A}^a)_{\mu\nu}
-(i/2){\epsilon^a}_{bc}({\cal A}^b
 \wedge {\cal A}^c)_{\mu\nu}$
is the curvature 2-form.

The equations of motion for the dynamical variables
(${\cal A}^a_i$ and $\tilde{E}^i_a$) are
\begin{eqnarray} \partial_t {\cal A}^a_{i} &=&
-i \epsilon^{ab}_{~~c}\null \! \mathop {\vphantom {N}\smash N}
\limits ^{}_{^\sim}\!\null \tilde{E}^j_{b} F_{ij}^{c}
+N^j F^a_{ji} +{\cal D}_i{\cal A}^a_{0}+e\Lambda \null \!
\mathop {\vphantom {N}\smash N}\limits ^{}_{^\sim}\!\null e^a_i,
\label{eqA} \\
\partial_t {\tilde{E}^i_a}
&=&-i{\cal D}_j( \epsilon^{cb}_{~~a} \null \!
\mathop {\vphantom {N}\smash N}\limits ^{}_{^\sim}\!\null
\tilde{E}^j_{c}
\tilde{E}^i_{b})
+2{\cal D}_j(N^{[j}\tilde{E}^{i]}_{a})
+i{\cal A}^b_{0} \epsilon_{ab}^{~~c} \tilde{E}^i_c,  \label{eqE}
\end{eqnarray}
\noindent
where
${\cal D}_jX^{ji}_a:=\partial_jX^{ji}_a-i
 \epsilon_{ab}^{~~c} {\cal A}^b_{j}X^{ji}_c,$
 for $X^{ij}_a+X^{ji}_a=0$.


In order to construct metric variables from the variables
$({\cal A}^a_i, \tilde{E}^i_a, \null \!
\mathop {\vphantom {N}\smash N}\limits ^{}_{^\sim}\!\null, N^i)$,
we first prepare
tetrad $E^\mu_I$ as
$E^\mu_{0}=({1 / e \null \! \mathop {\vphantom {N}\smash N}
\limits ^{}_{^\sim}\!\null}, -{N^i / e \null \!
\mathop {\vphantom {N}\smash N}\limits ^{}_{^\sim}\!\null})$ and
$E^\mu_{a}=(0, \tilde{E}^i_{a} /e).$
Using them, we obtain metric $g^{\mu\nu}$ such that
\begin{equation}
g^{\mu\nu}:=E^\mu_{I} E^\nu_{J} \eta^{IJ}. \label{recmet}
\end{equation}

Notice that in general the metric (\ref{recmet}) is not real.
To ensure the metric is real-valued,
we need to impose real lapse and shift vectors together with
two conditions (metric reality condition);
\begin{eqnarray}
{\rm Im} (\tilde{E}^i_a \tilde{E}^{ja} ) &=& 0, \label{w-reality1} \\
{\rm Re} (\epsilon^{abc}
\tilde{E}^k_a \tilde{E}^{(i}_b {\cal D}_k \tilde{E}^{j)}_c)
&=& 0,
\label{w-reality2-final}
\end{eqnarray}
where the latter comes from the secondary condition of reality
of the metric
${\rm Im} \{ \partial_t(\tilde{E}^i_a \tilde{E}^{ja} ) \} = 0$
\cite{AshtekarRomanoTate}, and
we assume
${\rm det}\tilde{E}>0$ (see \cite{ys-con}).

For later convenience, we also prepare
stronger reality conditions.
These conditions are
\begin{eqnarray}
{\rm Im} (\tilde{E}^i_a ) &=& 0
\label{s-reality1} \\
{\rm and~~}
{\rm Im}  ( \partial_t {\tilde{E}^i_a} ) &=& 0,
\label{s-reality2}
\end{eqnarray}
\noindent
and we call them the ``primary triad reality condition" and the
``secondary triad
reality condition", respectively.
Using the equations of motion of $\tilde{E}^i_{a}$,
the gauge constraint (\ref{c-g}),
the metric reality conditions 
(\ref{w-reality1}), (\ref{w-reality2-final})
and the primary condition (\ref{s-reality1}),
we see  that  (\ref{s-reality2}) is equivalent to \cite{ys-con}
\begin{equation}
{\rm Re}({\cal A}^a_{0})=
\partial_i( \null \! \mathop {\vphantom {N}\smash N}
\limits ^{}_{^\sim}\!\null )\tilde{E}^{ia}
+(1 /2e) e^b_i\null \! \mathop {\vphantom {N}\smash N}
\limits ^{}_{^\sim}\!\null\tilde{E}^{ja}\partial_j\tilde{E}^i_b
+N^{i}{\rm Re}({\cal A}^a_i), \label{s-reality2-final}
\end{equation}
or with un-densitized variables,
\begin{equation}
{\rm Re}({\cal A}^a_{0})=
\partial_i( N)
{E}^{ia}
+N^{i}{\rm Re}({\cal A}^a_i).
\label{s-reality2-final2}
\end{equation} From 
this expression we see that
the second triad reality condition
restricts the three components of ``triad lapse" vector
${\cal A}^a_{0}$.
Therefore (\ref{s-reality2-final}) is
not a restriction on the dynamical variables
(${\cal A}^a_i$ and $\tilde{E}^i_a $)
but on the slicing, which we should impose on each hypersurface.
Thus the second triad reality condition does not restrict the
dynamical variables any
further than the second metric condition does.

\section{Hyperbolic formulation} \label{sec3}

We start from defining hyperbolic system following
Friedrichs \cite{Friedrichs54}, which is first applied in general
relativity by Fischer and Marsden \cite{FischerMarsden72}.
That is, we say that the system is first-order (quasi-linear) 
hyperbolic if a certain pair of variables $u_i $ form a linear system as
\begin{equation}
\partial_t u_i
=A^{lj}_{~~i}(u) \partial_l u_j 
+B_i(u),
\label{def}
\end{equation}
where $A$ is a characteristic matrix-valued function,
of which eigenvalues are all real, 
and $B$ is a function. 
We further define that the system is
symmetric when $A$ is a Hermitian matrix \cite{fried96,CH}.

The symmetric system gives us the energy 
integral inequalities,
which are the primary tools for analizing well-posedness of the system.
As was discussed by Geroch \cite{Geroch}, most physical systems
are expressed as symmetric hyperbolic systems.

Ashtekar's formulation itself is in the first-order
form in the sence of (\ref{def}),
but not a symmetric hyperbolic form. 

We start from writing the
principal part of the Ashtekar's evolution equations
as
\begin{equation}
\partial_t \left[ \begin{array}{l}
\tilde{E}^i_a \\
{\cal A}^a_i
\end{array} \right] \cong
 \left[ \begin{array}{cc}  A^{l~bi}_{~a~~j} & B^{l~~ij}_{~ab} \\
C^{lab}_{~~~ij} & D^{la~~j}_{~~bi}
\end{array} \right]
\partial_l
\left[ \begin{array}{l}
\tilde{E}^j_b \\
{\cal A}^b_j
\end{array} \right]
\label{matrixform}
\end{equation}
where
$\cong$ means that we extracted only the terms which
appear in the principal part of the system.
The system is symmetric hyperbolic if
\begin{eqnarray}
0&=&
A^{labij}-\bar{A}^{lbaji},  \label{cond1}
\\
0&=&
D^{labij}
-\bar{D}^{lbaij}, \label{cond2}
\\
0 &=& 
B^{labij}-\bar{C}^{lbaji}. \label{cond3}
\end{eqnarray}
where bar denotes complex conjugate.\footnote{We think that
the reader will not confuse $A^{labij}$ and $B^{labij}$ with
matrix $A$ and $B$  in (\ref{def}).}

We first prepare the constraints (\ref{c-ham})-(\ref{c-g}) as
\begin{eqnarray}
{\cal C}_{H} &\cong&
 i\epsilon^{ab}_{~~c} \tilde{E}^i_a \tilde{E}^j_b 
\partial_i{\cal A}^c_j
=
 i\epsilon^{dc}_{~~b} \tilde{E}^l_d \tilde{E}^j_c 
(\partial_l{\cal A}^b_j)
\nonumber \\
&=&
 -i\epsilon_b^{~cd} \tilde{E}^j_c \tilde{E}^l_d
(\partial_l{\cal A}^b_j),
\\
{\cal C}_{M k} &=&
  -F^a_{kj} \tilde{E}^j_a
\cong
  -(\partial_k{\cal A}^a_j-\partial_j {\cal A}^a_k) \tilde{E}^j_a
\nonumber \\
&=&
[-\delta^l_k \tilde{E}^j_b
+\delta^j_k \tilde{E}^l_b
](\partial_l{\cal A}^b_j),
\\
{\cal C}_{Ga} &=&
 {\cal D}_i \tilde{E}^i_a
\cong
\partial_l\tilde{E}^l_a.
\end{eqnarray}

We apply the same technique with ILR to modify the
equation of motion of $\tilde{E}^i_a$ and ${\cal A}^a_i$ by
adding the constraints which weakly produce
${\cal C}_{H} = 0,
{\cal C}_{M k} = 0$,  and
${\cal C}_{Ga} = 0$.
With a parametrization for triad lapse ${\cal A}^a_0$
with $T$ and $S$ as
\begin{equation}
\partial_i{\cal A}^a_0 \cong
 T_{~i~j}^{l~a~b} \partial_l \tilde{E}^j_b
+S_{~i~~b}^{l~aj} \partial_l {\cal A}^b_j, \label{A0katei}
\end{equation}
we write the principal parts of (\ref{eqA}) and (\ref{eqE}) as
\begin{eqnarray}
\partial_t \tilde{E}^i_a
&=&
-i{\cal D}_j( \epsilon^{cb}_{~~a} 
\null \! \mathop {\vphantom {N}\smash N}\limits ^{}_{^\sim}\!\null
\tilde{E}^j_c\tilde{E}^i_b)
+2{\cal D}_j(N^{[j}\tilde{E}^{i]}_a)
\nonumber \\&&
+i{\cal A}^b_0 \epsilon_{ab}^{~~c} \tilde{E}^i_c
+P^i_{~ab} {\cal C}_G^b
\nonumber\\&\cong&
-i\epsilon^{cb}_{~~a} 
\null \! \mathop {\vphantom {N}\smash N}\limits ^{}_{^\sim}\!\null 
(\partial_j\tilde{E}^j_c)\tilde{E}^i_b
-i\epsilon^{cb}_{~~a} 
\null \! \mathop {\vphantom {N}\smash N}\limits ^{}_{^\sim}\!\null 
\tilde{E}^j_c(\partial_j\tilde{E}^i_b)
\nonumber \\&&
+{\cal D}_j(N^j\tilde{E}^i_a)
-{\cal D}_j(N^i\tilde{E}^j_a)
+P^{i~b}_{~a} \partial_j\tilde{E}^j_b
\nonumber\\&\cong&
[-i\epsilon^{bc}_{~~a} 
\null \! \mathop {\vphantom {N}\smash N}\limits ^{}_{^\sim}\!\null 
\delta^l_j  \tilde{E}^i_c
-i\epsilon^{cb}_{~~a} 
\null \! \mathop {\vphantom {N}\smash N}\limits ^{}_{^\sim}\!\null 
\tilde{E}^l_c \delta^i_j
\nonumber \\&&
+N^l\delta^i_j \delta^b_a
-N^i\delta^l_j \delta^b_a
+P^{i~b}_{~a} \delta^l_j ]
(\partial_l\tilde{E}^j_b), \label{eqE2}
\end{eqnarray}
\begin{eqnarray}
\partial_t {\cal A}^a_i &=&
-i \epsilon^{ab}_{~~c}
\null \! \mathop {\vphantom {N}\smash N}\limits ^{}_{^\sim}\!\null 
\tilde{E}^j_b F^c_{ij}
+N^j F^a_{ji}
\nonumber \\&&
+{\cal D}_i{\cal A}^a_0+e\Lambda 
\null \! \mathop {\vphantom {N}\smash N}\limits ^{}_{^\sim}\!\null 
e^a_i
+Q^a_i {\cal C}_H+R_i^{~ja} {\cal C}_{Mj}
\nonumber\\&\cong&
-i \epsilon^{ab}_{~~c}
\null \! \mathop {\vphantom {N}\smash N}\limits ^{}_{^\sim}\!\null 
\tilde{E}^j_b (\partial_i{\cal A}^c_j-\partial_j{\cal A}^c_i)
+N^j (\partial_j{\cal A}^a_i-\partial_i{\cal A}^a_j)
\nonumber \\&&
+T_{~i~j}^{l~a~b} \partial_l \tilde{E}^j_b
+S_{~i~~b}^{l~aj} \partial_l {\cal A}^b_j
-Q^a_i i\epsilon_b^{~cd} 
\tilde{E}^j_c \tilde{E}^l_d  (\partial_l{\cal A}^b_j)
\nonumber \\&&
+R_i^{~ka} [-\delta^l_k \tilde{E}^j_b
+\delta^j_k \tilde{E}^l_b ]
\partial_l{\cal A}^b_j
\nonumber\\&\cong&
[
+i \epsilon^{a~c}_{~b}
\null \! \mathop {\vphantom {N}\smash N}\limits ^{}_{^\sim}\!\null 
\tilde{E}^j_c \delta^l_i
-i \epsilon^{a~c}_{~b}
\null \! \mathop {\vphantom {N}\smash N}\limits ^{}_{^\sim}\!\null 
\tilde{E}^l_c \delta^j_i
+N^l \delta^a_b \delta^j_i
-N^j \delta^a_b \delta^l_i
\nonumber \\&&
+S_{~i~~b}^{l~aj}
-iQ^a_i \epsilon_b^{~cd} \tilde{E}^j_c \tilde{E}^l_d
-R_i^{~la} \tilde{E}^j_b
+R_i^{~ja} \tilde{E}^l_b
](\partial_l{\cal A}^b_j)
\nonumber \\&&
+T_{~i~j}^{l~a~b} \partial_l \tilde{E}^j_b, \label{eqA2}
\end{eqnarray}
where $P, Q$ and $R$ are parameters and will be
fixed later.
Note that
we truncated ${\cal A}^a_0$ in (\ref{eqE2}),
while it remains in  (\ref{eqA2}), since
only the derivative of  ${\cal A}^a_0$ effects the principal
part of the system. From these two equations, we get
\begin{eqnarray}
A^{labij}&=&
-i\epsilon^{bca} 
\null \! \mathop {\vphantom {N}\smash N}\limits ^{}_{^\sim}\!\null 
\gamma^{lj}\tilde{E}^i_c
-i\epsilon^{cba} 
\null \! \mathop {\vphantom {N}\smash N}\limits ^{}_{^\sim}\!\null 
\tilde{E}^l_c \gamma^{ij}
\nonumber \\&&
+N^l\gamma^{ij} \delta^{ab}
-N^i\gamma^{lj} \delta^{ab}
+P^{iab} \gamma^{lj},
\\
B^{labij}&=&0,
\\
C^{labij}&=&
T^{liajb},
\\
D^{labij}&=&
+i \epsilon^{abc}
\null \! \mathop {\vphantom {N}\smash N}\limits ^{}_{^\sim}\!\null 
\tilde{E}^j_c \gamma^{li}
-i \epsilon^{abc}
\null \! \mathop {\vphantom {N}\smash N}\limits ^{}_{^\sim}\!\null 
\tilde{E}^l_c \gamma^{ji}
\nonumber \\&&
+N^l \delta^{ab} \gamma^{ji}
-N^j \delta^{ab} \gamma^{li}
+S^{liajb}
\nonumber \\&&
-iQ^{ai} \epsilon^{bcd} \tilde{E}^j_c \tilde{E}^l_d
-R^{ila} \tilde{E}^{jb}
+R^{ija} \tilde{E}^{lb}.
\end{eqnarray}

The condition (\ref{cond1}) is
written as
\begin{eqnarray}
0&=&
-i\epsilon^{bca} 
\null \! \mathop {\vphantom {N}\smash N}\limits ^{}_{^\sim}\!\null 
\gamma^{lj}\tilde{E}^i_c
-i\epsilon^{acb} 
\null \! \mathop {\vphantom {N}\smash N}\limits ^{}_{^\sim}\!\null 
\gamma^{li}\bar{\tilde{E}}^j_c
\nonumber \\&&
-2i\epsilon^{cba} 
\null \! \mathop {\vphantom {N}\smash N}\limits ^{}_{^\sim}\!\null 
{\rm Im}(\tilde{E}^l_c) \gamma^{ij}
-N^i\gamma^{lj} \delta^{ab}
\nonumber \\&&
+N^j\gamma^{li} \delta^{ba}
+P^{iab} \gamma^{lj}
-\bar{P}^{jba} \gamma^{li}. \label{a-a}
\end{eqnarray}
Because 
the third term in the right-hand-side
cannot be elliminated using $P$, we
assume the triad reality condition 
${\rm Im}(\tilde{E}^l_c) =0$
hereafter. 
Then (\ref{cond2}) and (\ref{cond3}) become
\begin{eqnarray}
0
&=&\bar{T}^{ljbia} \label{b-c}
\\
0
&=&
i \epsilon^{abc}
\null \! \mathop {\vphantom {N}\smash N}\limits ^{}_{^\sim}\!\null 
\tilde{E}^j_c \gamma^{li}
+i \epsilon^{bac}
\null \! \mathop {\vphantom {N}\smash N}\limits ^{}_{^\sim}\!\null 
\tilde{E}^i_c \gamma^{lj}
-N^j \delta^{ab} \gamma^{li}
+N^i \delta^{ba} \gamma^{lj}
\nonumber \\&&
+S^{liajb}-\bar{S}^{ljbia}
-iQ^{ai} \epsilon^{bcd} \tilde{E}^j_c \tilde{E}^l_d
-i\bar{Q}^{bj} \epsilon^{acd} \tilde{E}^i_c \tilde{E}^l_d
\nonumber \\&&
-R^{ila} \tilde{E}^{jb}
+R^{ija} \tilde{E}^{lb}
+\bar{R}^{jlb} \tilde{E}^{ia}
-\bar{R}^{jib} \tilde{E}^{la}.  \label{d-d}
\end{eqnarray}
The third and forth term in (\ref{d-d}) cannot be elliminated
using $Q$ or $R$, so
 $S^{liajb}=\gamma^{li}\delta^{ab}N^j$ is determined.
Thus $S$ and ${T}^{ljbia}=0$ [eq. (\ref{b-c})]
decides the form of the triad lapse as
\begin{equation}
{\cal A}^a_0 = {\cal A}^a_jN^j+\mbox{non-dynamical~terms}
\label{shift-triad-relation}
\end{equation}
in result. In order to be consistent with the triad lapse condition
(\ref{s-reality2-final2}), 
we need to specify the lapse as $\partial_i N=0$.
This lapse condition is also supported by the fact that if we do not
assume $\partial_i N=0$, then   the secondary triad reality condition 
(\ref{s-reality2-final}) makes the 
system second order. 
ILR does not discuss consistency of the system with 
reality condition (especially with secondary reality condition). 
However, since ILR assume ${\cal A}^a_0 = {\cal A}^a_jN^j$,
we think that ILR also needs to impose similar restricted lapse
condition in order to preserve reality of the system.

The rest of our effort is finished when we specify parameters 
$P$, $Q$ and $R$.
$P$ is given by decomposing (\ref{a-a}) into real/complex parts;
\begin{eqnarray}
0
&=&
-N^i\gamma^{lj} \delta^{ab}
+N^j\gamma^{li} \delta^{ba}
\nonumber \\&&
+{\rm Re}(P)^{iab} \gamma^{lj}
-{\rm Re}(P)^{jba} \gamma^{li}
\\
0
&=&
-\epsilon^{bca} 
\null \! \mathop {\vphantom {N}\smash N}\limits ^{}_{^\sim}\!\null 
\gamma^{lj}\tilde{E}^i_c
-\epsilon^{acb} 
\null \! \mathop {\vphantom {N}\smash N}\limits ^{}_{^\sim}\!\null 
\gamma^{li}\tilde{E}^j_c
\nonumber \\&&
+{\rm Im}(P)^{iab} \gamma^{lj}
+{\rm Im}(P)^{jba} \gamma^{li}
\end{eqnarray}
By multiplying $\gamma_{li}$ in these two and taking symmetric and
anti-symmetric operation to the index $ab$, we obtain
\begin{equation}
P^{iab}=
N^i \delta^{ab}+i
\null \! \mathop {\vphantom {N}\smash N}\limits ^{}_{^\sim}\!\null 
\epsilon^{abc}\tilde{E}^i_c.
\end{equation}
For $Q$ and $R$, we found that a combination of the choice
\begin{eqnarray}
Q^{ai}&=&
e^{-2}
\null \! \mathop {\vphantom {N}\smash N}\limits ^{}_{^\sim}\!\null 
\tilde{E}^{ia}
\\
R^{ila}&=&
i e^{-2}
\null \! \mathop {\vphantom {N}\smash N}\limits ^{}_{^\sim}\!\null 
\epsilon^{acd} \tilde{E}^i_d \tilde{E}^l_c
\end{eqnarray}
satisfies the condition (\ref{d-d}).

\section{Discussion} \label{sec:disc}
In summary, by adding constraint terms with 
appropriate coefficients,
we succeed to construct a symmetric hyperbolic formulation for the
Ashtekar's system. This formulation is consistent with secondary triad
reality condition, which requires to impose a constant lapse function
for the evolving system.

The characteristic speeds of this system are given 
by finding eigenvalues
of the characteristic matrix $A$ of (\ref{def}).
Since $A$ is a Hermitian, eigenvalues of $A$ are all real.
Then it is again clear that this system is symmetric hyperbolic. 
Actually the eigenvalues of the $18 \times 18$ matrix $A^l$
for $x^l$-direction are:
$ N^l $ (multiplicity = 6),
$ N^l \pm \sqrt{\gamma^{ll}} N $
 (5 each),
 and
$ N^l \pm 3\sqrt{\gamma^{ll}} N$
(1 each), where we do not take the sum in 
$\gamma^{ll}$ here.
These speeds are independent from the way of taking a triad.
We omit to show the related eigen-vectors because of
saving space.

As we denoted in \S 3, 
our formulation requires triad reality condition. 
In order to make the system first order, the lapse function 
is assumed to be constant. 
Shift vectors and triad lapse ${\cal A}^a_0$ should have a relation
(\ref{shift-triad-relation}). This can be interpreted that shift 
is free and triad 
lapse is determined. 
This gauge restriction sounds tight, but this 
arises from our general assumption of (\ref{A0katei}). 
There might be a possibility to improve the situation 
by renormalizing shift and triad lapse terms into 
left-hand-side of
equations of motion like the case of 
GR \cite{ChoquetBruhatYork}.
Or this might be 
because our 
system is constituted by Ashtekar's original 
variables. 
We are now trying to release this gauge restriction and/or
to simplify the characteristic speeds
by other gauge possiblities and also 
by introducing new dynamical variables. 
This effort will be reported elsewhere.


\vspace{0.4cm}

This work (HS) was partially supported by NSF PHYS 96-00049, 96-00507,
and NASA NCCS 5-153.


\end{document}